\documentclass[twocolumn,english,aps,prl,amsmath,amssymb,nofootinbib,showpacs,superscriptaddress]{revtex4}
\usepackage[latin9]{inputenc}
\usepackage{bm}
\usepackage{amsmath}
\usepackage{amssymb}
\usepackage{graphicx}
\usepackage{esint}

\makeatletter
\@ifundefined{textcolor}{}
{%
 \definecolor{BLACK}{gray}{0}
 \definecolor{WHITE}{gray}{1}
 \definecolor{RED}{rgb}{1,0,0}
 \definecolor{GREEN}{rgb}{0,1,0}
 \definecolor{BLUE}{rgb}{0,0,1}
 \definecolor{CYAN}{cmyk}{1,0,0,0}
 \definecolor{MAGENTA}{cmyk}{0,1,0,0}
 \definecolor{YELLOW}{cmyk}{0,0,1,0}
}

\usepackage[english]{babel}
\usepackage{latexsym}
\usepackage{graphics}
\usepackage{epsfig}
\usepackage{color}
\usepackage{bm}

\makeatother

\usepackage{babel}
\begin{document}

\title{Traveling Majorana solitons in a one-dimensional spin-orbit coupled
Fermi superfluid}

\author{Peng Zou}

\affiliation{Centre for Quantum and Optical Science, Swinburne University of Technology,
Melbourne 3122, Australia}

\author{Joachim Brand}

\affiliation{New Zealand Institute for Advanced Study, Centre of Theoretical Chemistry
and Physics, \\
 and Dodd-Walls Centre for Photonic and Quantum Technologies, Massey
University, Auckland, New Zealand }

\author{Xia-Ji Liu}

\affiliation{Centre for Quantum and Optical Science, Swinburne University of Technology,
Melbourne 3122, Australia}

\author{Hui Hu}

\email{hhu@swin.edu.au}

\affiliation{Centre for Quantum and Optical Science, Swinburne University of Technology,
Melbourne 3122, Australia}

\date{\today}
\begin{abstract}
We investigate traveling solitons of a one-dimensional spin-orbit
coupled Fermi superfluid in both topologically trivial and non-trivial
regimes by solving the static and time-dependent Bogoliubov-de Gennes
equations. We find a critical velocity $v_{h}$ for traveling solitons
that is much smaller than the value predicted using the Landau criterion
due to the presence of spin-orbit coupling, which strongly upshifts
the energy level of the soliton-induced Andreev bound states towards
the quasi-particle scattering continuum. Above $v_{h}$, our time-dependent
simulations in harmonic traps indicate that traveling solitons decay
by radiating sound waves. In the topological phase, we predict the
existence of peculiar Majorana solitons, which host two Majorana fermions
and feature a phase jump of $\pi$ across the soliton, irrespective
of the velocity of travel. These unusual properties of Majorana solitons
may open an alternative way to manipulate Majorana fermions for fault-tolerant
topological quantum computations. 
\end{abstract}


\maketitle
Solitons or localized waves that arise from the interplay between
the dispersion and nonlinearity of underlying systems are fascinating
phenomena occurring in many different fields of physics \cite{SolitonIntroduction}.
Over the past two decades, a major research emphasis has focused on
solitons in atomic Bose-Einstein condensates (BECs) \cite{SolitonBECs}.
The family of BEC solitons consists of many interesting members, from
bright solitons in attractive BECs \cite{Denschlag2000} and gap solitons
in optical lattices \cite{Ostrovskaya2003}, to dark solitons in repulsively
interacting BECs \cite{Burger1999,Anderson2001,Becker2008}, which
are created experimentally by imprinting a sharp and characteristic
phase jump into the BEC. Remarkably, dark solitons may also be created
in strongly interacting Fermi gases \cite{Antezza2007,Scott2011,Spuntarelli2011,Liao2011,Scott2012,Yefsah2013,Bulgac2014,Ku2014,Efimkin2015}
at the crossover from BECs to Bardeen-Cooper-Schrieffer (BCS) superfluids
\cite{Giorgini2008}, where phase kinks are encoded in the pairing
order parameter. Their recent experimental observation may offer valuable
insights into the nature of fermionic superfluidity in the strongly
correlated regime \cite{Yefsah2013,Ku2015}.

\begin{figure}
\centering \includegraphics[scale=0.6]{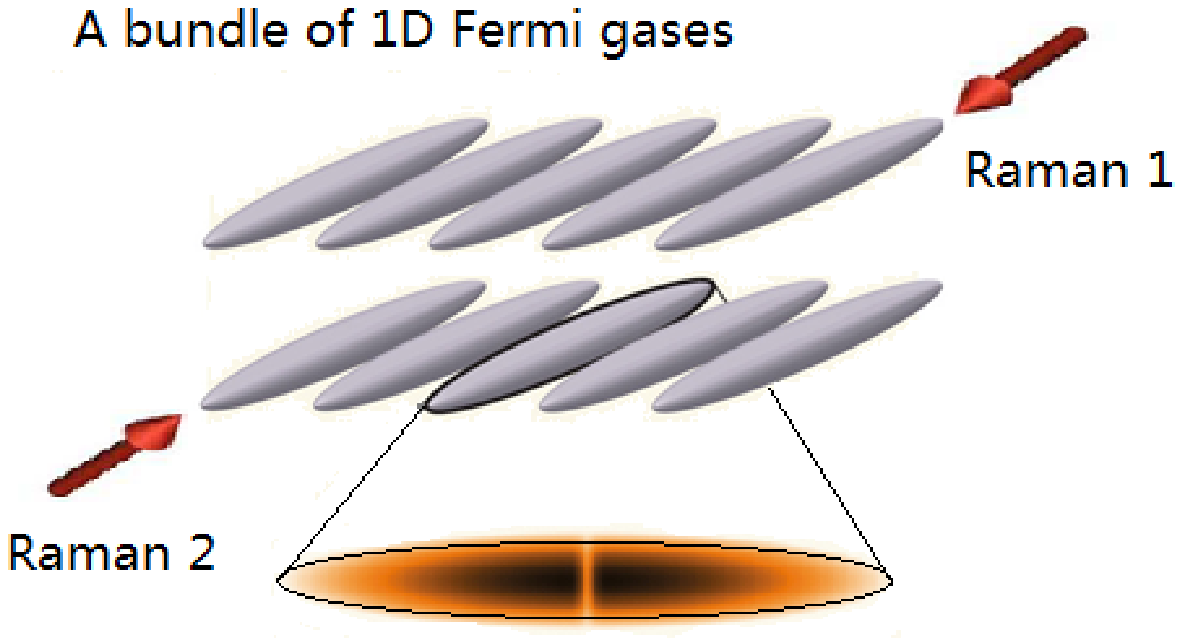}

\includegraphics[scale=0.43]{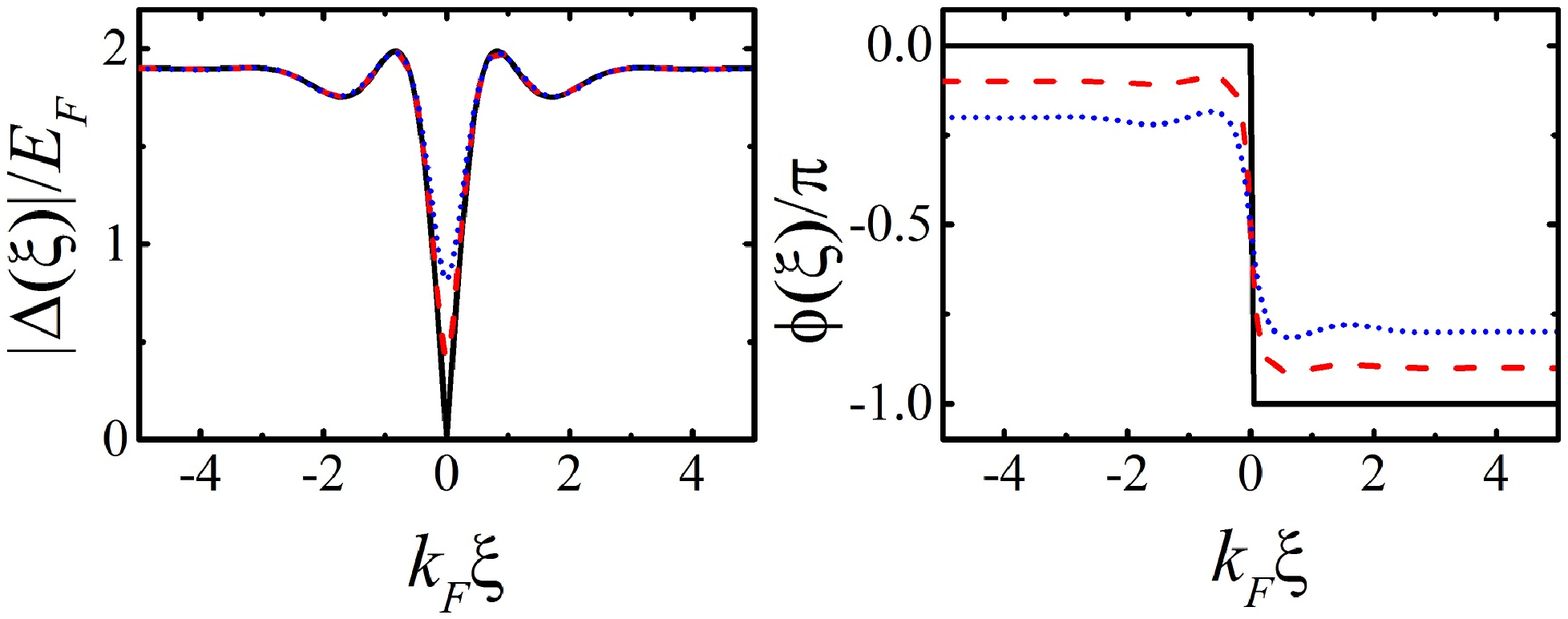} \protect\caption{(color online). Upper panel: Sketch of the proposed experimental configuration.
Many 1D tubes of spin-orbit coupled $^{40}$K Fermi gases are formed
using a 2D optical lattice and two counter-propagating Raman lasers.
Lower panel: The magnitude $|\Delta(\xi)|$ and phase $\phi(\xi)$
of the soliton order parameter in the non-topological phase with parameters
$\gamma=3.41$, $h=0.52E_{F}$ and $\lambda k_{F}/E_{F}=1.71$. The
solid, dashed and dotted lines correspond to soliton velocities $v_{s}=0$,
$0.15v_{F}$ and $0.3v_{F}$, respectively. }

\label{fig1} 
\end{figure}

In this Letter, we consider traveling fermionic solitons in a different
setup -- one-dimensional (1D) Fermi superfluids with spin-orbit coupling
(see Fig.~\ref{fig1}) -- and predict the existence of an exotic
member of the soliton family when the superfluid becomes topologically
non-trivial. It is referred to as Majorana soliton, owing to its ability
to host two Majorana fermions that obey non-Abelian statistics at
the soliton core \cite{Majorana1937,Wilczek2009}. Majorana solitons
are universal and remarkably robust, in the sense that their properties
are not affected by a finite velocity of travel. In particular, the
phase jump across a Majorana soliton is exactly pinned to $\pi$ and
the density profile is unchanged. In other words, Majorana solitons
are not greyed by a finite velocity. This unique stability renders
Majorana solitons an ideal platform to manipulate Majorana fermions
for practical applications such as topological quantum computations
\cite{Nayak2008}.

Our investigation is motivated by the recent realizations of spin-orbit
coupling in atomic Fermi gases \cite{Wang2012,Cheuk2012} and the
promising perspective of creating an atomic topological superfluid
\cite{Liu2012,Wei2012}. Traveling Majorana solitons with fixed $\pi$
phase step, if experimentally observed to oscillate inside a Fermi
cloud, would be a smoking-gun proof of the existence of long-sought
topological superfluids. We note that stationary dark solitons with
Majorana fermions in a spin-orbit coupled Fermi gas were recently
predicted \cite{Xu2014,Liu2015}. However, the crucial issue raised
in any practical manipulations, i.e., the fate of these solitons at
a finite velocity of motion, was not addressed.

Our results also suggest that a critical velocity for the stability
of traveling solitons is greatly affected by spin-orbit coupling.
This is because one of the mid-gap energy levels, the soliton-induced
Andreev bound states (ABSs), is strongly up-shifted towards the bulk
quasi-particle scattering continuum by spin-orbit coupling. As a result,
the Andreev bound states are lost at a much smaller velocity $v_{h}$
than the Landau critical velocity, which leads to a decay channel.
At a velocity above $v_{h}$, we find that traveling solitons gradually
decay via radiating sound waves.

\textit{Model.} We start by describing a possible experimental configuration,
as sketched in the upper panel of Fig.~\ref{fig1}. A bundle of parallel,
identical 1D spin-1/2 $^{40}$K Fermi gases can be formed by adding
a tight 2D optical lattice in the transverse $y-z$ plane \cite{Liao2010},
and the spin-orbit coupling with equal Rashba and Dresselhaus weight
can be realized by adapting the so-called NIST scheme using two counter-propagating
Raman laser beams \cite{Wang2012}. The resulting 1D spin-orbit coupled
Fermi gas in a single tube is modeled by the Hamiltonian $H=\int dx\left[\mathcal{H}_{0}+\mathcal{H}_{{\rm {int}}}\right]$,
where \cite{Liu2012,Wei2012,Xu2014,Liu2015} 
\begin{equation}
\mathcal{H}_{0}=\left[\psi_{\uparrow}^{\dagger}\left(x\right),\psi_{\downarrow}^{\dagger}\left(x\right)\right]\left(\mathcal{H}_{s}+\lambda\hat{k}_{x}\sigma_{y}-h\sigma_{z}\right)\left[\begin{array}{c}
\psi_{\uparrow}\left(x\right)\\
\psi_{\downarrow}\left(x\right)
\end{array}\right]
\end{equation}
is the spin-orbit coupled single-particle part and 
\begin{equation}
\mathcal{H}_{{\rm int}}=g_{{\rm 1D}}\psi_{\uparrow}^{\dagger}\left(x\right)\psi_{\downarrow}^{\dagger}\left(x\right)\psi_{\downarrow}\left(x\right)\psi_{\uparrow}\left(x\right)
\end{equation}
with $g_{1D}<0$ is the interaction Hamiltonian describing the attractive
contact interaction between the two spin states ($\sigma=\uparrow,\downarrow$).
Here, $\psi_{\sigma}^{\dagger}$ is the fermionic field operator that
creates an atom with mass $m$ in the spin state $\sigma$. The term
$\lambda\hat{k}_{x}\sigma_{y}-h\sigma_{z}$ with the momentum operator
$\hat{k}_{x}=-i\partial/\partial x$ and Pauli matrices $\sigma_{y}$
and $\sigma_{z}$ is induced by the Raman process, describing a synthetic
spin-orbit coupling with strength $\lambda\equiv\hbar^{2}k_{R}/m$
and an effective Zeeman field $h=\Omega_{R}/2$, where $k_{R}$ and
$\Omega_{R}$ are the momentum and Rabi frequency of the Raman beams
\cite{Wang2012}, respectively. The term $\mathcal{H}_{s}=-\hbar^{2}\partial_{x}^{2}/(2m)+V_{T}(x)-\mu$
with the chemical potential $\mu$ describes the motion of atoms in
a harmonic trapping potential $V_{T}(x)=m\omega^{2}x^{2}/2$.

We solve the model Hamiltonian for stationary and traveling solitons
within the mean-field approximation. This amounts to finding solutions
with \emph{phase-twisted} order parameter in the static and time-dependent
Bogoliubov-de Gennes (BdG) equations, $\mathcal{H}_{{\rm BdG}}\bm{\Phi}_{\eta}(x)=E_{\eta}\bm{\Phi}_{\eta}(x)$
and $\mathcal{H}_{{\rm BdG}}\bm{\Phi}_{\eta}(x,t)=i\hbar(\partial/\partial t)\bm{\Phi}_{\eta}(x,t)$,
respectively. Here, for convenience we have used the Nambu spinor
representation and have introduced $\mathbf{\Phi}_{\eta}\equiv[u_{\uparrow\eta},u_{\downarrow\eta},v_{\uparrow\eta},v_{\downarrow\eta}]^{T}$
and $E_{\eta}$ as the wave-function and energy of Bogoliubov quasiparticles.
The BdG Hamiltonian reads 
\begin{equation}
\mathcal{H}_{{\rm BdG}}\equiv\begin{bmatrix}\mathcal{H}_{s}-h & -\lambda\partial/\partial x & 0 & -\Delta\\
\lambda\partial/\partial x & \mathcal{H}_{s}+h & \Delta & 0\\
0 & \Delta^{*} & -\mathcal{H}_{s}+h & \lambda\partial/\partial x\\
-\Delta^{*} & 0 & -\lambda\partial/\partial x & -\mathcal{H}_{s}-h
\end{bmatrix},\label{HBdG}
\end{equation}
and the BdG equations, either static or time-dependent, should be
self-consistently solved with the gap equation $\Delta=-(g_{1D}/2)\sum_{\eta}[u_{\uparrow\eta}v_{\downarrow\eta}^{*}f(E_{\eta})+u_{\downarrow\eta}v_{\uparrow\eta}^{*}f(-E_{\eta})]$
and the number equation $n=-(1/2)\sum_{\eta}[|u_{\sigma\eta}|^{2}f(E_{\eta})+|v_{\sigma\eta}|^{2}f(-E_{\eta})]$,
where $f(E)=1/(1+e^{{\it E/{\it k_{B}T}}})$ is the Fermi-Dirac distribution
function and the summation is performed for the energy level (labeled
by $\eta$) up to a high-energy cut-off $E_{c}$, i.e., $\left|E_{\eta}\right|<E_{c}$.

To obtain a moving soliton in a trapped gas, we first find a stationary
dark soliton at $x_{0}$ away from the trap center \cite{Liu2015}.
By evolving such an initial state in time, the soliton is accelerated
by the trap potential and caused to oscillate inside the Fermi cloud.
The same procedure has previously been used to understand the dynamics
of dark solitons in a BEC-BCS Fermi superfluid \cite{Scott2011},
and could also be employed in experiment.

We also search for traveling soliton solutions on a homogeneous (untrapped)
background that satisfy $\Delta(x,t)=\Delta(x-v_{s}t)=\Delta(\xi)$,
by solving the BdG equations in the co-moving frame with the velocity
$v_{s}$ \cite{Liao2011}:
\begin{equation}
\mathcal{H}_{{\rm BdG}}(\xi)\bm{\Phi}_{\eta}(\xi)=\left[E_{\eta}-i\hbar v_{s}\frac{\partial}{\partial\xi}\right]\bm{\Phi}_{\eta}(\xi).\label{vBdG}
\end{equation}
Here, $\mathcal{H}_{{\rm BdG}}(\xi)$ is obtained by replacing $\partial_{x}$
with $\partial_{\xi}$ and $\Delta(x,t)$ with $\Delta(\xi)$ in Eq.~(\ref{HBdG}).
In other words, we seek traveling solitons in a homogeneous gas that
are stationary in the frame of the soliton. This technique provides
more insights into the soliton properties and enables us to isolate
effects caused by the trapping potential when we analyze time-dependent
simulations \cite{Scott2012}. For the calcuations in a box with length
$L$ we impose a modified periodic boundary condition, $\Delta(\xi+L/2)=\Delta(\xi-L/2)e^{i\delta\phi}$,
to explicitly take into account a phase jump $\delta\phi$ across
the soliton \cite{Efimkin2015}. In addition, we implement a generalized
secant (Broyden's) approach to make sure that the self-consistent
iteration procedure will converge to a stable solution \cite{Liao2011,Baran2008}.

In numerical calculations, we use a dimensionless interaction parameter
to characterize the interaction strength, $\gamma=-mg_{1D}/(\hbar^{2}n)$,
which is basically the ratio between the interaction and kinetic energy
at the density $n$. We choose the Fermi vector and energy, $k_{F}=\pi n/2$
and $E_{F}=\hbar^{2}k_{F}^{2}/(2m)$, as the units of wave-vector
and energy, respectively. For simulations in a trapped cloud with
$N$ atoms, it is convenient to use the peak density of a \emph{non-interacting}
Fermi gas in the Thomas-Fermi approximation at the trap center, $n'=(2/\pi)\sqrt{Nm\omega/\hbar}$,
although the cloud itself is an interacting gas. We denote the corresponding
units with $k_{F}'$ and $E_{F}'$. Throughout this work, we consider
only zero temperature. For trapped simulations we shall take the interaction
parameter $\gamma\simeq3$, spin-orbit coupling strength $\lambda k_{F}'/E_{F}'=1.5$
and an energy cut-off $E_{c}=10E_{F}'$. Parameters for homogeneous
simulations are chosen to correspond to the relevant peak density
of the interacting trapped gas. 

There are two different regimes for a 1D spin-orbit-coupled Fermi
superfluid \cite{Liu2012,Wei2012}, depending on whether the effective
Zeeman field $h$ is over a threshold $h_{c}=\sqrt{\Delta^{2}+\mu^{2}}$
($\simeq E_{F}$ with our parameters for the trapped cloud). Once
$h>h_{c}$, the superfluid becomes topologically non-trivial and hosts
Majorana solitons. Before presenting our main results on Majorana
solitons, it is useful to understand how traveling solitons are affected
by spin-orbit coupling in the non-topological phase.

\begin{figure}
\centering \includegraphics[scale=0.43]{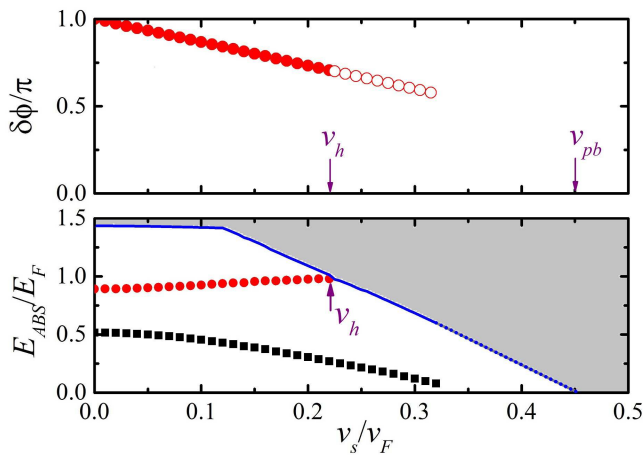} \protect\caption{(color online). Upper panel: The phase jump $\delta\phi$ as a function
of velocity in the non-topological phase. Lower panel: The corresponding
mid-gap ABS energy levels. The arrows indicate the velocity $v_{h}$,
at which the upper ABS (red circles) touches the quasi-particle scattering
continuum (shadow area), and the pair-breaking velocity $v_{pb}$.
Broyden's method fails to find traveling soliton solutions when the
energy of the lower ABS is close to zero. Parameters are as in Fig.~\ref{fig1}.}

\label{fig2} 
\end{figure}

\textit{Non-topological phase}. The spatial structure of the soliton
order parameter in the nontopological phase ($h<h_{c}$) is illustrated
in the lower panel of Fig.~\ref{fig1} for different soliton velocities.
As the velocity increases, the dip in the order parameter profile
becomes shallower, and its imaginary part develops structure and becomes
larger at the soliton core. Consequently, the phase jump across the
soliton decreases from $\pi$, as shown explicitly in the upper panel
of Fig.~\ref{fig2}. This turn-to-grey procedure of traveling solitons
has been predicted earlier both for BEC-BCS crossover superfluids
\cite{Liao2011} and BECs \cite{Tsuzuki1971,Konotop2004}. However,
the presence of spin-orbit coupling leads to some interesting new
features. 

\begin{figure}
\centering \includegraphics[scale=0.43]{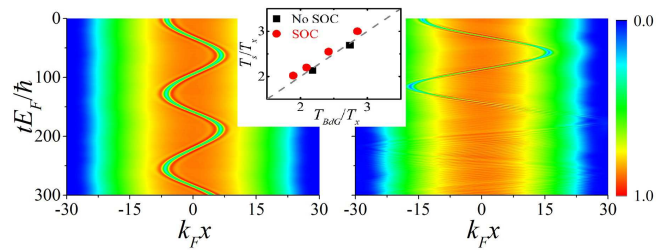} \protect\caption{(color online). Time-dependent simulations of traveling solitons in
a trapped, non-topological Fermi superfluid with $h=0.4E_{F}'<h_{c}$.
The color represents the magnitude of density (in units of $n'$).
By choosing the initial position $x_{0}$, we generate two solitons,
whose maximum velocity is $0.18v_{F}'<v_{h}=0.22v_{F}'$ (left panel)
and $0.42v_{F}'>v_{h}$ (right panel), respectively. The inset examines
the universal relation (\ref{eq:TsTx}) for the soliton oscillation
period with/without spin-orbit coupling at different interaction strengths
($2.5\leq\gamma\leq3.2$). The period $T_{s}$ from the time-dependent
simulation is compared with $T_{BdG}$ defined by the right hand side
of Eq. (\ref{eq:TsTx}) and calculated from the time-independent BdG
solutions.}

\label{fig3} 
\end{figure}

The most striking feature is that the mid-gap energy levels of soliton-induced
ABSs now exhibit a pronounced velocity dependence, as seen from the
lower panel of Fig.~\ref{fig2}. Already at zero velocity, the ABS
splits into two branches due to the combined effects of spin-orbit
coupling and effective Zeeman field \cite{Xu2014,Liu2015}. With increasing
the soliton velocity, the energy of the upper ABS gradually increases
and merges into the quasi-particle scattering continuum at $v_{h}\simeq0.22v_{F}$,
which is much smaller than the pair-breaking velocity $v_{pb}\simeq0.45v_{F}.$
Any coupling between the upper ABS and the bulk continuum states \cite{footnote1}
then will destroy the soliton-induced ABSs and in turn make the soliton
unstable. Thus, we anticipate that the soliton may decay when its
velocity is beyond the threshold $v_{h}$, for example, by dissipating
its energy in the form of sound waves.

\begin{figure}
\centering \includegraphics[scale=0.43]{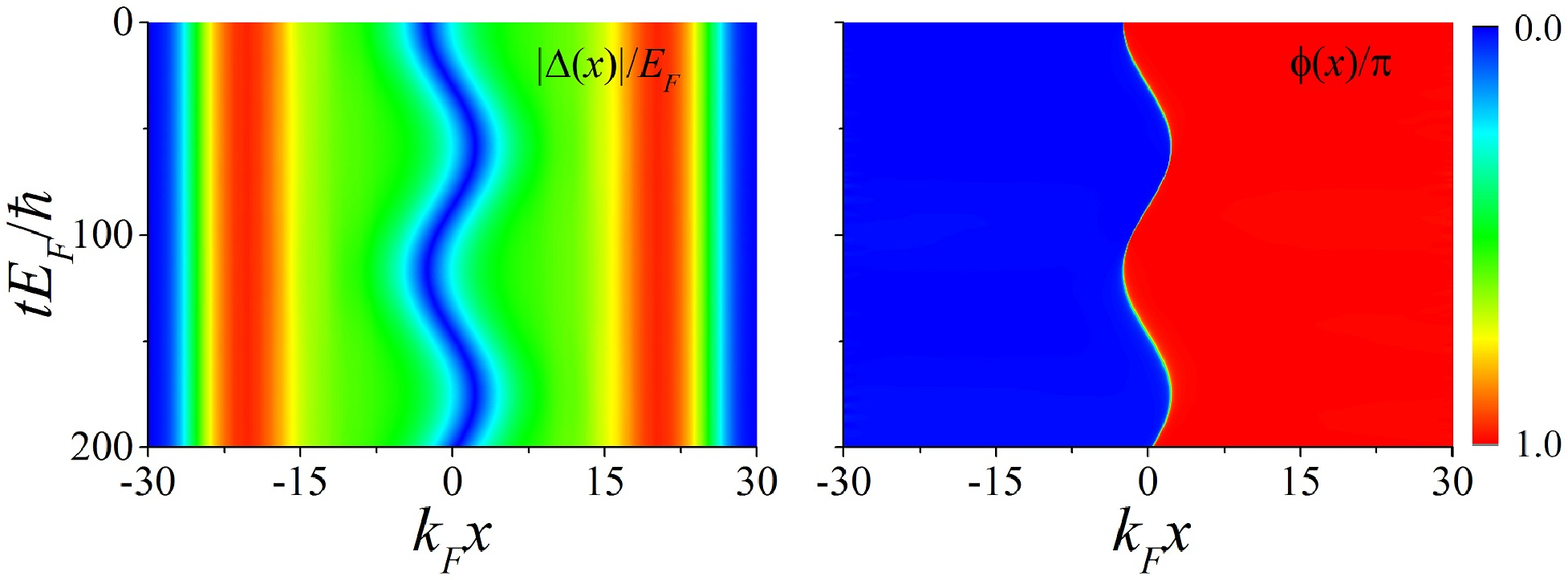}

\includegraphics[scale=0.43]{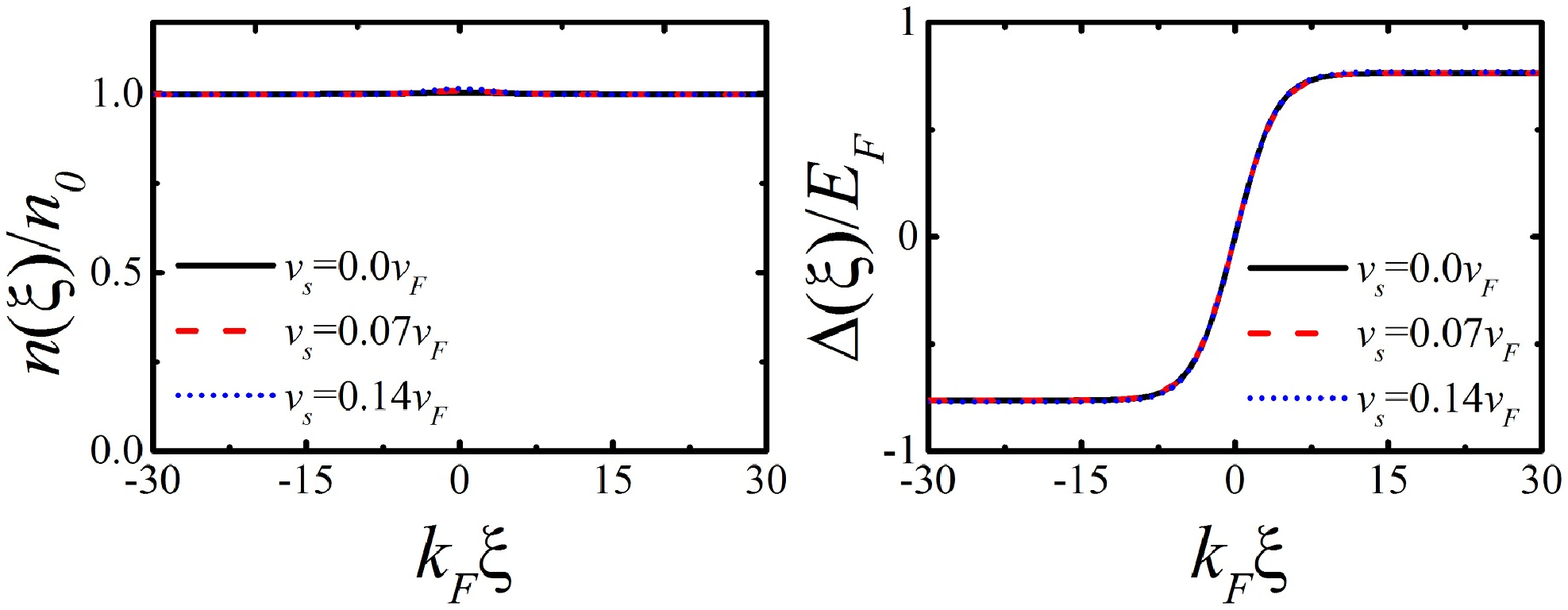} \protect\caption{(color online). Majorana soliton. The two upper panels report the
time evolution of the magnitude $\left|\Delta(x,t)\right|$ (left)
and the phase $\phi(x,t)$ (right) of the order parameter for a Majorana
soliton in a trapped topological Fermi superfluid ($h=1.2E_{F}'>h_{c}$)
with a maximum soliton velocity $0.07v_{F}<v_{h}$. The two lower
panels show the density (left) and order parameter (right) of a Majorana
soliton in the homogeneous configuration at different velocities with
parameters $\gamma=3.75$, $h=1.71E_{F}$ and $\lambda k_{F}/E_{F}=1.79$. }

\label{fig4} 
\end{figure}

We have checked this conjecture by performing time-dependent simulations
in harmonic traps, as reported in Fig.~\ref{fig3}. By carefully
selecting the position $x_{0}$ of the initially stationary dark soliton,
the maximum velocity $v_{m}$ - reached when the traveling soliton
passes the trap center - can be tuned. For $v_{m}<v_{h}$, we find
a stable oscillation of the traveling soliton (see the left panel).
The oscillation period $T_{s}$ seems to satisfy the elegant universal
relation (see the inset),
\begin{equation}
\left(\frac{T_{s}}{T_{x}}\right)^{2}=\frac{M^{*}}{M}=1+\frac{\hbar n}{2M}\frac{d\left(\delta\phi\right)}{dv_{s}},\label{eq:TsTx}
\end{equation}
which was derived by treating soliton as a classical particle \cite{Scott2011,Konotop2004}.
Here, $T_{x}=2\pi/\omega$ is the trapping period, $M$ and $M^{*}$
are respectively the physical and inertial mass of the soliton and
their difference is proportional to the derivative of the phase jump
\cite{Scott2011}. In contrast, at $v_{m}>v_{h}$, the soliton gradually
spreads out in the density profile and after a few periods we see
only low-amplitude density ripples (right panel). By examining the
speed of these ripples, we identify them as sound waves. Our time-dependent
simulations with spin-orbit coupling therefore indicate that the critical
velocity of traveling solitons could be significantly smaller than
Landau critical velocity $v_{pb}$, which was found to be the relevant
critical velocity without spin-orbit coupling \cite{Scott2011,Spuntarelli2011,Liao2011}. These
result are still consistent, since without spin-orbit coupling $v_{h}$
actually is close to the pair-breaking velocity \cite{Scott2012}.

\begin{figure}
\centering \includegraphics[scale=0.43]{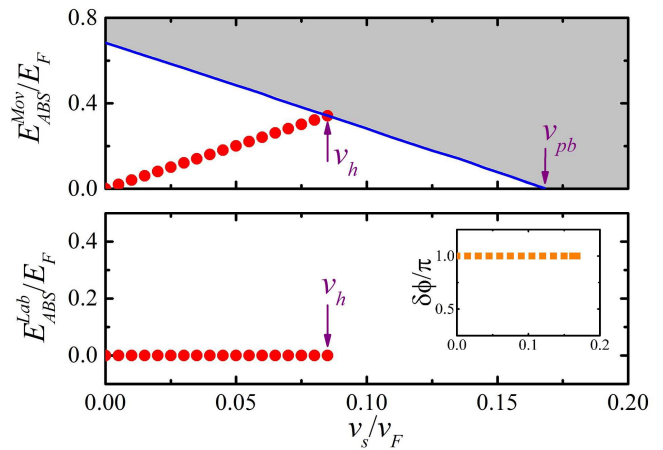} \protect\caption{(color online). The ABS energy of the Majorana soliton as a function
of the soliton velocity in the co-moving frame (upper panel) or in
the laboratory frame (lower panel). We note that in the topological
phase, the number of the ABS states decreases to one \cite{Liu2015},
if we count only positive energy levels. The inset examines the $\pi$-phase
of Majorana solitons. Parameters as in Fig.~\ref{fig4}.}

\label{fig5} 
\end{figure}

\textit{Topological phase}. By increasing effective Zeeman field across
$h_{c}\simeq E_{F}$ for a trapped Fermi cloud, the local energy gap
(and hence the pair-breaking velocity) at the trap center closes and
then re-opens. A topological superfluid emerges.The first sign of
the existence of a velocity-independent Majorana soliton comes from
the time-dependent simulations in harmonic traps, as shown in the
upper panel of Fig.~\ref{fig4}. During the time evolution, the dip
minimum in $\left|\Delta(x,t)\right|$ remains at zero and the phase
jump $\delta\phi(t)$ across the soliton is always pinned at $\pi$
(see also the inset in Fig.~\ref{fig5}). In the lower panel of Fig.~\ref{fig4},
we check more rigorously the velocity dependence using Broyden's approach.
With increasing the soliton velocity in the topological phase, the
density and pairing order parameter profiles remain essentially unchanged.

To show the presence of Majorana fermions at the soliton core, we
report in Fig.~5 the energy of the ABS as a function of the traveling
velocity. Although in the co-moving frame the energy $E_{ABS}^{Mov}$
increases (linearly) with the velocity, the energy in the laboratory
frame, $E_{ABS}^{\mathrm{Lab}}$, which is related to the co-moving
energy by 
\begin{equation}
E_{ABS}^{\mathrm{Lab}}=E_{ABS}^{\mathrm{Mov}}+\int d\xi\mathbf{\Phi}_{ABS}^{*}(-i\hbar v_{s})\partial\mathbf{\Phi}_{ABS}/\partial\xi,
\end{equation}
is precisely zero \cite{footnote2}. This is expected behavior for
a Majorana fermion, which must have zero energy due to the particle-antiparticle
symmetry. Together with the observed continuity with the zero velocity
case \cite{Liu2015,Xu2014}, we conclude that the moving soliton in
the topological phase indeed hosts Majorana fermions.The properties
of the Majorana soliton at finite velocity can be made plausible from
the universal relation (\ref{eq:TsTx}), if we assume its validity
in the topological phase. We recall that the density notch in Majorana
solitons is absent \cite{Xu2014,Liu2015} and hence the physical mass vanishes
\cite{Scott2011}. Equation (\ref{eq:TsTx}) immediately implies that
the derivative of the phase jump is zero, since the oscillation period
should be finite. This leads to a constant $\pi$ phase jump, irrespective
of the soliton velocity. In turn, the magnitude of order parameter
should vanish at the soliton core. 

It is worth noting that Eq.~(\ref{eq:TsTx}) can hardly be used to
predict the oscillation period of Majorana solitons, as the ratio
between zero mass and zero derivative of the phase jump is undetermined.
How to amend the universal relation for Majorana solitons needs further
exploration. For a large velocity, we find a similar situation as
in the non-topological case (see the upper panel of Fig.~\ref{fig5}).
Naïvely, we anticipate that Majorana solitons may cease to exist once
$v_{s}>v_{h}$. However, time-dependent simulations in traps suggest
that the instability via emitting sound waves occurs at a very long
time scale, presumably due to the weak coupling between the ABS and
quasi-particle continuum. This again indicates the robustness of Majorana
solitons.

\textit{Experimental observation of Majorana solitons}. The proposed
experimental scheme in Fig.~\ref{fig1} is easy to set up \cite{Wang2012,Liao2010},
although the realization of 1D topological superfluids is difficult
due to the lack of efficient cooling techniques \cite{Wang2012}.
Majorana solitons can be created by imprinting a sharp phase jump
\cite{Yefsah2013,Ku2014}. The observation of their oscillations seems
to be an experimental challenge, as the density profile of Majorana
solitons remains flat \cite{Xu2014,Liu2015}. One then has to measure
the local pairing order parameter. A suitable detection technique
is the spatially resolved radio-frequency spectroscopy \cite{Shin2007},
which may give information about the local order parameter, provided
that the size of Majorana solitons is comparable with the spatial
resolution of spectroscopy. 

\textit{Conclusions and outlook}. We have predicted the existence
of an exotic member to the soliton family - the Majorana soliton -
which may exist universally in any 1D topological superfluids including
$p$-wave superfluids and semiconductor/superconductor nanowire structures
\cite{Mourik2012}. Our results may also be applicable to 2D topological
superfluids. In that case, it would be interesting to examine the
possibility of finding a moving vortex that is able to host a single
Majorana fermion in the vortex core. In analogy to Majorana solitons,
the properties of such a vortex would be insensitive to its velocity.
\begin{acknowledgments}
We are grateful to Franco Dalfovo, Robin Scott and Yan-Hua Hou for
fruitful discussions. This research was supported by the Marsden Fund
of New Zealand (contract number UOO1320) and by the ARC Discovery
Projects (FT130100815, DP140100637, DP140103231 and FT140100003).\end{acknowledgments}


\begin{thebibliography}{10}
\bibitem{footnote1}Broyden's approach does not capture such a coupling
and can predict traveling soliton solutions at $v_{s}>v_{h}$.
\bibitem{footnote2}Due to the particle-hole symmetry, there is another
ABS branch whose energy decreases linearly with the traveling velocity
in the co-moving frame. In the lab frame, its energy is also precisely
zero, giving rise to another Majorana fermion hosted by the Majorana
soliton.
\end{thebibliography}
\end{document}